\documentclass[8pt]{article}
\usepackage{amssymb,amsmath,epsfig}
\allowdisplaybreaks

\begin{document}

\title{\bf First and Second Laws of Thermodynamics in Modified Ho\v{r}ava-Lifshitz $F(R)$ gravity}

\author{\bf{Abdul Jawad}\thanks {jawadab181@yahoo.com; abduljawad@ciitlahore.edu.pk}
~and Shamaila Rani\thanks {drshamailarani@ciitlahore.edu.pk}\\
Department of Mathematics, COMSATS Institute of\\ Information
Technology, Lahore-54000, Pakistan.\\ \bf{Davood Momeni}
\thanks{momeni-d@enu.kz; d.momeni@yahoo.com}\\ Eurasian International Center for
Theoretical Physics\\ and Department of General Theoretical
Physics,\\ Eurasian National University, Astana 010008,
Kazakhstan.\\
\bf{Faiza Gulshan}\thanks{fazi.gull@yahoo.com}\\
Department of Mathematics, Lahore Leads university,\\
Lahore-54590, Pakistan.\\
\bf{Ratbay Myrzakulov} \thanks{rmyrzakulov@gmail.com}\\
Eurasian International Center for Theoretical Physics\\ and
Department of General Theoretical Physics,\\ Eurasian National
University, Astana 010008, Kazakhstan.\\}

\date{}

\maketitle
\begin{abstract}

\end{abstract}
In this paper we discuss the thermodynamics of the apparent horizon
in $F(R)$ Ho\v{r}ava-Lifshitz gravity in equilibrium and
non-equilibrium ensembles. We show that the second law of thermodynamics can be
satisfied in this non relativistic theory. 
\section{Introduction}
Different types of the observational data, namely type Ia supernovae , cosmic microwave background (CMB) , large scale structure , baryon acoustic oscillations , and weak lensing show that our Universe is accelerating \cite{Ri98,PeRa03} . Modified gravity is the simplest way to address this accelerating behavior. In this approach one simply modifies the original Einstein-Hilbert action by an arbitrary function of the curvature term(s) like $R,R_{\mu\nu},R^{\alpha}_{\beta\mu\nu},...$. Such types of modifications originally proposed in \cite{Buchdahl} and recently revisited in light of the current acceleration of the Universe
 \cite{RevNoOd},\cite{Nojiri:2006ri}. The simplest model of modified gravity is a class of models, alled $F(R)$ gravity in which one replaces the classical Hilbert-Einstein action of gravity by an arbitrary function of $R$, the Ricci scalar term \cite{Carroll:2003wy}:
\begin{eqnarray}
S=\frac{1}{2\kappa^2}\int d^4x\sqrt{-g}F(R)
\end{eqnarray}
Different aspects of this type of gravity studied in literature
\cite{viablemodels}. One significant result is that
  the $F(R)$ gravity is Lorentz invariant like Einstein gravity because of its invariant form under global coordinate transformations. This feature is a basic concept and is valid even in the  teleparallel gravity \cite{Momeni:2015uwx}. Modified gravity looks very similar to the Einstein gravity in solar system when
 all solar tests by a highly precision are done, we observe that small deviations from this theory can satisfy these local tests. This is one of the most important advantages of $F(R)$ gravity
\cite{solartests,Olmo07} . Based on this fact, we are able to successfully reconstruct viable models of $f(R)$ gravity for cosmological applications
\cite{Hu:2007nk,solartests2,Sawicki:2007tf,Amendola:2007nt}. It was proven that the
$F(R)$ models are responsible for accelerating expansion as well as to describe the dark matter problem in the  rotation curves of different galaxies
without the need for  dark
matter . This issue is vastly studied by authors
\cite{Cap2,Borowiec:2006qr,Mar1,Boehmer:2007kx,Bohmer:2007fh}(see for example
 \cite{RevNoOd,
SoFa08}.) Thee are other types of modified gravity theories for example when the
Ricci scalar $R$
is coupled to the matter Lagrangian density $L_ m$ both in metric approach \cite{Bertolami:2007gv},
\cite{ha08}, \cite{Bertolami:2007vu, ha10} and in the
Palatini formulation
 \cite{Pal}.  Also another version of these non-minimally coupled models has been proposed
 in \cite{Rlm},where they are
assuming that the gravitational Lagrangian is given by an
arbitrary function of the Ricci scalar $R$ and of the matter
Lagrangian $L_m$  in the form of  $f(R,L_m)$ gravity, this model originally
was proposed in \cite{Poplawski:2006ey}.

 Non relativistic regimes of gravity is also important for example to improve the propagator of graviton in ultraviolent regime. In (2009) a new approach to the quantum gravity proposed by
  Ho\v{r}ava based on idea of Lifshitz in  quantum systems \cite{hor3,hor2,hor1,hor4}. The proposal is to take into account different space-time footing. On account of  this assumpion, the   Lorentz symmetry is broken consequently  theory renormalizable at quantum level. We'll review this proposal more techincally in Sec. (\ref{review}). The theory called as   Ho\v{r}ava-Lifshitz theory and widely studied in literature
\cite{Volovik:2009av}-\cite{
chen}. Ho\v{r}ava-Lifshitz cosmology is alse widely studied by authors
\cite{Calcagni:2009ar,Kiritsis:2009sh, odi, sari1, sari2}. In
particular, one can examine specific solution subclasses
\cite{Lu:2009em,Nastase:2009nk,Minamitsuji:2009ii}, the
perturbation spectrum \cite{cai,
Gao:2009bx,Chen:2009jr,Gao:2009ht,Wang:2009yz,Kobayashi:2009hh,
ding}, the gravitational wave production
\cite{Takahashi:2009wc,Koh:2009cy}, the matter bounce
\cite{Brandenberger:2009yt,Brandenberger:2009ic,Cai:2009in}, the
black hole properties \cite{Danielsson:2009gi,Kehagias:2009is,
Mann:2009yx,Bertoldi:2009vn,Castillo:2009ci,BottaCantcheff:2009mp,
ohta:cao}, the cosmic string solutions \cite{mom} the dark energy
phenomenology
\cite{Saridakis:2009bv,Wang:2009rw,Appignani:2009dy,Setare:2009vm},
the astrophysical phenomenology \cite{Kim:2009dq,Harko:2009qr},exact solutions to field equations \cite{Setare:2009sw}-\cite{Momeni:2009au}
, quantum spectrum of black holes  \cite{Setare:2010gi}
and
etc
\cite{Charmousis:2009tc,Sotiriou:2009bx,Bogdanos:2009uj}.\par
Notivated by $F(R)$ gravity a fully foliation-preserving diffeomorphisms invariance version of HL theory proposed in \cite{Chaichian:2010yi},\cite{Kluson:2009rk}. This viable extension of Ho\v{r}ava-Lifshitz theory has the following remarkable results:
\begin{itemize}
\item It was demontrated that cosmological equation on the spatially-flat  sector of space time are consistent with the constraint equations.
\item Due to  the existence of the de Sitter solutions in several versions of theory, it is possible to consistently unify  the early-time inflation with the late-time acceleration.
\item It reduces to the classical cosmological equationswith  a special choice of parameters.
\item The cosmological
equations do coincide with the ones for the related, convenient
$F(R)$ gravity. This means the cosmological history of
Ho\v{r}ava-Lifshitz $F(R)$ gravity will be just the same as for its
convenient version. For the general version of the theory the
situation turns out to be more complicated.

\end{itemize}
The cosmological viability of the model was demonstrated in $F(R)$-Ho\v{r}ava-Lifshitz theory but it is not east to construct  an appropriate consistent Hamiltonian formalism \cite{Kiriushcheva:2011xh},\cite{Kiriushcheva:2011qf}.
 $F(R)$-HL theory has many possible uses in the cosmology and has also been investigated as a potential valid modification of the original HL theory
\cite{LopezRevelles:2012cg}-\cite{SaezGomez:2010ex}.
However, although the cosmological effects of the $F(R)$ Ho\v{r}ava-Lifshitz  on the physical properties of Universe  was demonstrated over last years ago, little attention has been paid to the thermodynamics of an appropriate $F(R)$ Ho\v{r}ava-Lifshitz  model.  The present paper presents a set of criteria for investigating  thermodynamic laws. On the basis of these criteria it then describes the validation of a first and second laws  using apparent and event horizons.
This combination of two basically distinct laws formed a novel interpretation in which the incorporation of  $F(R)$ part  significantly increased  viability.

This paper is organized as follows: In Sec. (\ref{review}) we present a brief review to the original HL theory. In Sec. (\ref{F(R)HL}) we provide the basicl Eqs. in $F(R)$ Ho\v{r}ava-Lifshitz theory. In Sec. (\ref{cosmology}) we study cosmological solutions. Sec. (\ref{Thermo}) is devoted to study equilibrium and non-equilibrium thermodynamics. In Sec. (\ref{equil}) we study equilibrium regime. We summarize in final section.

\section{Review of Ho\v{r}ava-Lifshitz gravity with detailed condition}
\label{review}
Ho\v{r}ava-Lifshitz theory is
a power-counting renormalizable,ultraviolet
complete theory of gravity
\cite{hor2,hor1,hor3,hor4}. The fixed point of theory in infrared regime is Einstein gravity. In the  UV regime, HL  theory
has a fixed point with an anisotropic, Lifshitz scaling
between time and space of the form $x^{i}\to\ell~x^{i}$,
$t\to\ell^z~t$, where $\ell$, $z$, $x^{i}$ and $t$ are the scaling
factor, dynamical critical exponent, spatial coordination and
temporal coordination, respectively. Let us to start by decomposing metric in ADM formalism.
Following from the Arnowitt-Dese-Misner formalism (ADM) decomposition of the metric  \cite{ADM}.\cite{ADM2},\cite{ADM3},
  and the Einstein equations, the dynamical fields
are the fields $N(t,x),N_{i}(t,x),g_{ij}(t,x)$ corresponding to
the \emph{lapse }, \emph{shift} and \emph{spatial metric} .
The general metric in the so-called ADM
decomposition in a $3+1$ spacetime \cite{ADM}-\cite{ADM3} is
\begin{equation}\label{F1}
ds^{2}=-N^{2}dt^{2}+g_{ij}^{(3)}(dx^{i}+N^{i}dt)(dx^{j}+N^{j}dt),
\end{equation}
where $i,j=1,2,3$, $N$ is the so-called Lapse variable and $N^{i}$
is the shift $3$-vector. The Ricci scalar in the general relativity
$(GR)$ can be written in terms of the metric and we have
\begin{equation}\label{F2}
R=K_{ij}K^{ij}-K^{2}+R^{(3)}+2\nabla_{\mu}(n^{\mu}\nabla_{\nu}n^{\nu}-n^{\nu}\nabla_{\nu}n^{\mu}),
\end{equation}
here we define the extrinsic curvature  $K=g^{ij}K_{ij}$, $K_{ij}$, the spatial scalar curvature of $g_{ij}$ is denoted by
$R^{(3)}$  and $n^{\nu}$ is a unit normal
vector in con dimension one time sliced metric $t=0$. We can
define the extrinsic curvature as
\begin{equation}\label{F3}
K_{ij}=\frac{1}{2N}(\dot{g}_{ij}^{(3)}-\nabla_{i}^{(3)}N_{j}-\nabla_{j}^{(3)}N_{i}),
\end{equation}
the lapse variable $N$ in the original model is taken to be
just time-dependent and the condition of projectability is hold. Using the foliation-preserving diffeomorphisms invariance, we can fix  $N=1$. This version of theory is called projectable and it was demonstrated that may cause problems with Newton's
lawin weak limit \cite{Blas}. To preserve Newtonian gravity in weak regimes, we need to work in the framework of  the non-projectable $F(R)$-model \cite{Chaichain2}. For the non-projectable case, the Newton's
law could be restored by the "healthy" extension of the original
 Ho\v{r}ava gravity of \cite{Blas}.

The action of Ho$\check{\textbf{r}}$ava-Lifshitz theory  for
$z=3$ is
\begin{eqnarray}
S=\int_{M} dtd^{3}x\sqrt{g} N(\mathcal{L}_{K} - \mathcal{L}_{V} )
\end{eqnarray}
here the space-covariant derivative on a covector $v_{i}$ is defined by
$\nabla_{i}v_{j}\equiv \partial_{i}v_{j}-\Gamma_{ij}^{l}v_{l}$
where $\Gamma_{ij}^{l}$ are the spatial Christoffel symbols, by $g$ we mean
the determinant of the 3-metric $g_{ij}$ and $N = N(t)$ is a dimensionless
homogeneous gauge field. The kinetic term is
\begin{eqnarray}\nonumber
\mathcal{L}_{K}=\frac{2}{\kappa^2}\mathcal{O}_{K}=\frac{2}{\kappa^2}(K_{ij}K^{ij}-\lambda
K^2)
\end{eqnarray}
Here $N_{i} $ is a gauge field with scaling dimension $[N_{i}] =
z -
1$.\\
The \emph{potentialí} term $\mathcal{L}_{V}$ of the
$(3+1)$-dimensional theory is determined by the \emph{principle of
detailed balance } is given by the following expression:
\begin{eqnarray}
\mathcal{L}_{V}=\alpha_{6}C_{ij}C^{ij} -
\alpha_{5}\epsilon_{l}^{ij} R_{im}\nabla_{j}R^{ml} + \alpha_{4}
[R_{ij}R^{ij}- \frac{4\lambda-1}{4(3\lambda-1)} R^2]
+\alpha_{2}(R - 3\Lambda_{W})
\end{eqnarray}
The coupling constants $\alpha_{i}$ define by
\begin{eqnarray}\nonumber
\alpha_{2}=\frac{\alpha_{4}\Lambda_{w}}{3\lambda-1},\ \
\alpha_{4}=\frac{\kappa^2\mu^2}{8},\ \
\alpha_{6}=\frac{\kappa^2}{2\nu^4},\ \ \alpha_{5}=\frac{\kappa^2\mu}{2\nu^2}
\end{eqnarray}

 Where in it $C_{ij}$
is the \emph{Cotton }tensor  which is defined as,
\begin{eqnarray}\nonumber
C^{ij}=\epsilon^{kl(i}\nabla_{k}R^{j)}_{l}
\end{eqnarray}
Following \cite{Lu:2009em} we can write the action as
\begin{eqnarray}
&&S=\int dtdx^3(\mathcal{L}_{0}+\mathcal{L}_{1})\\&&
\mathcal{L}_{0}=\sqrt{g}N(\frac{2}{\kappa^2}(K_{ij}K^{ij}-\lambda
K^2)+\frac{\kappa^2\mu^2(\Lambda_{w}R-3\Lambda_{w}^2)}{8(1-3\lambda)})\\&&
\mathcal{L}_{1}=\sqrt{g}N(\frac{\kappa^2\mu^2(1-4\lambda)}{32(1-3\lambda)}R^2-\frac{\kappa^2}{2w^4}(C^{ij}-\frac{\mu
w^2}{2}R^{ij})(C_{ij}-\frac{\mu w^2}{2}R_{ij}))
\end{eqnarray}
 A curious case in the  Ho\v{r}ava-Lifshitz theory is that the
new mode  satisfies a first order (in time derivatives)
equation of motion . In linear aproximation this extra freedom degree
manifested only around non-static spatially inhomogeneous
backgrounds. Blas. et. al \cite{Blas} modified HL theory because of the following  serious problems associated with this
mode.
\begin{itemize}

\item The mode develops very fast exponential
instabilities at short distances.

\item  It becomes strongly
coupled at an extremely low cutoff scale
\end{itemize}

The significantresult is that, so far it is proven that  Ho\v{r}ava-Lifshitz theory has a vacuum solution as LIfshitz metric. So, it enjoys holography principle in non relativistic regime \cite{Griffin:2012qx}.

\section{Modified $F(R)$  Ho\v{r}ava-Lifshitz Gravity}\label{F(R)HL}

The action for standard $F(R)$ gravity can be written as\cite{Chaichain2}
\begin{equation}\label{F4}
S=\int{d^{4}x\sqrt{g^{(3)}}NF(R)}.
\end{equation}
 If
we take $z=1$, then GR is recovered. We can rewwrite the action as follows:
\begin{eqnarray}\nonumber
S&=&\frac{1}{2\kappa^{2}}\int{dtd^{3}x\sqrt{g^{(3)}}NF(\tilde{R})},
\tilde{R}=K_{ij}K^{ij}-\lambda{K}^{2}\\\label{F7}&+&R^{(3)}+2\mu
\nabla_{\mu}(n^{\mu}\nabla{_{\nu}}n^{\nu}-n^{\nu}\nabla{_{\nu}}
n^{\mu})-L^{(3)}(g_{ij}^{(3)}),
\end{eqnarray}
where $\kappa$ is the dimensionless gravitational coupling, $\mu$
and $\lambda$ are the new constants which account for the violation
of full diffeomorphism transformations. Note that
the third term  in the expression for $\tilde{R}$ in the original
Horava gravity theory  can be omitted, as it becomes a total
derivative. The term $L^{(3)}(g_{ij}^{(3)})$ is written as
\begin{equation}\label{F8}
L^{(3)}(g_{ij}^{(3)})=E^{ij}G_{ijkl}E^{kl},
\end{equation}
where $G_{ijkl}$ is the inverse of the generalized De Witt metric
is,
\begin{equation}\label{F9}
G^{ijkl}=\frac{1}{2}(g^{(3)ik}g^{(3)jl}+g^{(3)il}g^{(3)jk})-{\lambda}g^{(3)ij}g^{(3)kl}.
\end{equation}
So, we have
\begin{equation}\label{F10}
G_{ijkl}=\frac{1}{2}(g_{ik}^{(3)}g_{jl}^{(3)}+g_{il}^{(3)}g_{jk}^{(3)})-
\bar{\lambda}g_{ij}^{(3)}g_{kl}^{(3)},
\bar{\lambda}=\frac{\lambda}{3\lambda-1}.
\end{equation}
Here it is important to note that $G^{ijkl}$ is singular for
$\lambda=1/3$ and $G_{ijkl}$ exist if $\lambda\neq1/3$.

  The expression for $E_{ij}$ is constructed to satisfy the
"detailed balance principle" \cite{6} and defined as
\begin{equation}\label{1}
\sqrt{g^{(3)}}E^{ij}=\frac{{\delta}W[g_{kl}^{(3)}]}{{\delta}g_{ij}^{(3)}},
\end{equation}
where the form of $W[g_{kl}^{(3)}]$ is given \cite{12} for $z=2$ and
$z=3$.

\section{Cosmology of $F(R)$ Ho\v{r}ava-Lifshitz theory}\label{cosmology}

We want to study of cosmological solutions for the theory described
 by action (\ref{F7}). The
spatially-flat  Friedmann-Lemaitre-Robertson-Walker (FLRW) metric is assumed as
\begin{equation}\label{2}
ds^{2}=-N^{2}dt^{2}+a^{2}(t){\sum}_{i=1}^{3}(dx^{i})^{2}.
\end{equation}
We can see that, $N$ can be taken to be just time-dependent in
projectability condition and can be fixed to be unity, $N=1$ by
using the foliation-preserving diffeomorphisms (\ref{F6}). $N$
depends on both time and spatial coordinates for the
non-projectability condition. So, the assumption of the solution $N$
is taken as unity.

For the metric (\ref{2}), the scalar $\tilde{R}$ is given by
\begin{equation}\label{3}
\tilde{R}=\frac{3(1-3\lambda+6\mu)H^{2}}{N^{2}}+\frac{6\mu}{N}\frac{d}{dt}(\frac{H}{N}).
\end{equation}
For the action (\ref{F7}), and assuming the FRW metric (\ref{3}),
the second FRW equation can be obtained by varying the action with
respect to the spatial metric $g_{ij}^{(3)}$, we get
\begin{eqnarray}\nonumber
0&=&F(\tilde{R})-2(1-3\lambda+3\mu)(\dot{H}+3H^{2})F'(\tilde{R})-
2(1-3\lambda)\\\label{4}&\times&\dot{\tilde{R}}F''(\tilde{R})+
2\mu(\dot{\tilde{R}}^{2}F^{(3)}(\tilde{R})+\ddot{\tilde{R}}
F''(\tilde{R}))-{\kappa}^{2}\rho_{m},
\end{eqnarray}
where $\kappa^{2}=16{\pi}G$, is the pressure of perfect fluid that
fills the universe, and $N=1$. Note that, this equation becomes the
usual second FLRW equation for convenient $F(\tilde R)$ gravity
(\ref{F4}) and the constants  $\lambda$, $\mu$ can be taken as
$\lambda=\mu=1$. If we take the projectability condition, then the
variation over $N$ of the action (\ref{F7}) can be written as global
constraint.
\begin{equation}\label{5}
0=\int{d^{3}x[F(\tilde{R})-6(1-3\lambda+3\mu)H^{2}-6{\mu}\dot{H}+
6{\mu}H\dot{\tilde{R}}F''(\tilde{R})-{\kappa}^{2}\rho_{m}]}.
\end{equation}
By using the ordinary conservation equation for the matter fluid
$\dot\rho_{m}+3H(\rho_{m}+p_{m})=0$ and by integrating Eq. (14), we
have
\begin{eqnarray}\nonumber
0&=&F(\tilde{R})-6[(1-3\lambda+3\mu)H^{2}-{\mu}\dot{H}]F'(\tilde{R})
+6{\mu}H\dot{\tilde{R}}\\\label{6}&\times&F''(\tilde{R})-{\kappa}^{2}
\rho_{m}-\frac{C}{a^{3}},
\end{eqnarray}
where $C$ is the integrating constant, taken to be zero, according
to the constraint equation (\ref{5}). On the other hand, if we take
the non-projectability condition, we can obtain the equation
(\ref{6}) directly which corresponds to the first FLRW equation, by
variation over $N$.

The scalar curvature (\ref{3}) can be written as
\begin{equation}\label{7}
\tilde{R}=3(1-3\lambda+6\mu)H^{2}+6{\mu}HH'.
\end{equation}

\section{Non-Equilibrium Description of Thermodynamics in
$F(R)$ Ho\v{r}ava-Lifshitz Gravity}\label{Thermo}
\subsection{Energy Density and Pressure of Dark Components}
The energy density and dark components can be evaluated by field
equations. So we can rewrite the FLRW Eqs. (\ref{4}) and (\ref{6}) in
$F(R)$ Ho\v{r}ava-Lifshitz gravity as
\begin{equation}\label{8}
H^{2}=\frac{\kappa^{2}}{3(3\lambda-1)}\rho_{_{eff}},\quad
\dot{H}=\frac{-\kappa^{2}}{2(3\lambda-1)}(\rho_{_{eff}}+p_{_{eff}}),
\end{equation}
where $H=\dot{a}/a$ is the Hubble parameter and dot denotes the
derivative w.r.t 't'.
\begin{equation}\label{9}
\rho_{_{eff}}=\hat\rho_{_{de}}+\rho_{_{m}}.
\end{equation}
\begin{equation}\label{10}
P_{_{eff}}=\hat p_{_{de}}+p_{_{m}}.
\end{equation}
here$\hat\rho_{_{de}}$ and $\hat P_{_{de}}$ are the energy density
and pressure of dark components are given by
\begin{eqnarray}\nonumber
\hat\rho_{_{de}}&=&\frac{1}{\kappa^{2}}[-f(\tilde{R})+{3}
(1-3\lambda+6\mu)H^{2}F(\tilde{R})+{6}{\mu}{\dot{H}F
(\tilde{R})}\\\label{11}&-&{6}{\mu}H\dot{\tilde{R}}F'(\tilde{R})].
\end{eqnarray}
\begin{eqnarray}\nonumber
\hat p_{_{de}}&=&\frac{1}{\kappa^{2}}[f(\tilde{R})-{6}{\mu}
{\dot{H}}F(\tilde{{R}})-{3}(1-3\lambda+6\mu)H^{2}F(\tilde{R})
-2\\\label{12}&\times&(1-3\lambda)H{\dot{\tilde{R}}}F'
(\tilde{R})+{2}{\mu}{\dot{\tilde{R}}}^2F''(\tilde{R})+
{2}{\mu}{\ddot{\tilde{R}}F'(\tilde{R})}].
\end{eqnarray}

Leading to
\begin{equation}\label{13}
\dot{\hat\rho}_{_{de}}+3H(\hat\rho_{_{de}}+\hat
p_{_{de}})=-3(1-3\lambda)H^{2}\dot{F}.
\end{equation}
\begin{equation}\label{14}
\dot\rho_{_{m}}+3H\rho_{_{m}}=0.
\end{equation}
where, by hat we label all the  quantities in the non-equilibrium description of
thermodynamics . It is easy to check that the standard continuity equation does not hold due
to $\dot{F}\neq0$ in Eq. (\ref{13}), and dot denotes the derivative
with respect to 't' and prime denote the derivative with respect to
$\tilde R$.
\subsection{First Law of Thermodynamics}
In $F(R)$ Ho\v{r}ava-Lifshitz gravity by using the relation
$h^{\alpha\beta}\partial_{_{\alpha}}\tilde{r}\partial_{\beta}\tilde{r}=0$
we determines the dynamical apparent horizon. In the
flat FLRW spacetime, the radius $\tilde{r}_{_{A}}$ of the apparent
horizon is,
\begin{equation}\label{15}
\tilde r_{_{A}}=\frac{1}{H},
\end{equation}
the time derivative of Eq. (\ref{15}) is
\begin{equation}\label{16}
-\frac{d\tilde r_{_{A}}}{\tilde r_{_{A}}^{3}}=\frac{\dot H}{H}dt.
\end{equation}
Substituting Eq. (\ref{8}) into Eq. (\ref{16}) we get,
\begin{equation}\label{17}
\frac{F}{4\pi G}d\tilde r_{_{A}}=\frac{\tilde
r_{_{A}}^{3}H}{3\lambda-1}(\hat\rho_{_{t}}+\hat p_{_{t}})dt.
\end{equation}
Where $\hat\rho_{_{t}}\equiv\hat\rho_{_{de}}+\rho_{_{m}},\hat
p_{_{t}}\equiv\hat p_{_{de}}+p_{_{m}}$ are the total energy density
and pressure of the universe respectively.

The Bekenstein-Hawking horizon (killing) entropy is
defined as $S={A}/{4G}$, where $A=4\pi\tilde r_{_{A}}^{2}$ is the
area of the apparent horizon \cite{15}. In modified $F(\tilde R)$
gravity, a horizon entropy $\hat{S}$ associated with the Wald
entropy $\hat{S}$ is a Neother charge, is defined as
$\hat{S}={A}/4G_{_{eff}}$, where $G_{_{eff}}={G}/f'$ with
$f'=df(\tilde R)/d(\tilde R)$ is the effective gravitational
coupling in $F(\tilde R)$ gravity. It is remarkable to mention here that theWald
entropy $\hat{S}$ in $F(\tilde R)$ gravity kept the same form in both formalisms of
metric and palatini.

The entropy of black holes in $F(\tilde R)$ gravity is
\begin{equation}\label{18}
\hat{S}=\frac{FA}{4G}.
\end{equation}
By using Eqs. (\ref{17}) and (\ref{18}), we get
\begin{equation}\label{19}
\frac{1}{2\pi\tilde r_{_{A}}}d\hat S=\frac{4\pi}{3\lambda-1}
\tilde r_{_{A}}^{3}H(\hat\rho_{_{t}}
+\hat p_{_{t}})dt+\frac{\tilde r_{_{A}}}{2G}dF.
\end{equation}
The associated temperature of the apparent horizon has the following
Hawking temperature $T_{_{H}}$
\begin{equation}\label{20}
T_{_{H}}=\frac{|\kappa_{_{sg}}|}{2\pi}.
\end{equation}
Where $\kappa_{_{sg}}$ is the surface gravity
\begin{equation}\label{21}
\kappa_{_{sg}}=\frac{1}{2\sqrt{-h}}\partial_{_{0}}(\sqrt{-h}
h^{\alpha\beta}\partial_{_{\beta}}\tilde
r)
\end{equation}
\begin{eqnarray}\nonumber
\kappa_{_{sg}}&=&-\frac{1}{\tilde r_{_{A}}}(1-\frac{\dot{\tilde
r}_{_{A}}}{2H\tilde r_{_{A}}}) =-\frac{\tilde
r_{_{A}}}{2}(2H^{2}+\dot H)\\\label{22}&=&-\frac{2\pi
G}{3F(3\lambda-1)}(\hat\rho_{_{t}}-3\hat p_{_{t}}).
\end{eqnarray}
Where $h=det(h_{_{\alpha\beta}})$.
From Eq.(\ref{22}), we see that $\kappa_{_{sg}}\leq0$ if the total
equation of state (EoS)  $\omega_{_{t}}\equiv\hat
P_{_{t}}/\hat\rho_{_{t}}$ satisfies $\omega_{_{t}}\leq1/3$.

By solving the Eqs. (\ref{20}) and (\ref{22}), we have
\begin{equation}\label{23}
T_{_{H}}=\frac{1}{2\pi\tilde r_{_{A}}}(1-\frac{\dot{\tilde r}
_{_{A}}}{2H\tilde r_{_{A}}}).
\end{equation}
By multiplying the term $(1-\frac{\dot \tilde r_{_{A}}}{2H \tilde
r_{_{A}}})$ for Eq. (\ref{19}), we have
\begin{equation}\label{24}
T_{_{H}}d\hat{S}=\frac{4\pi}{3\lambda-1}\tilde
r_{_{A}}^{3}H(\hat\rho_{_{t}}+\hat
p_{_{t}})dt-\frac{2\pi}{3\lambda-1}\tilde
r_{_{A}}^{2}(\hat\rho_{_{t}}+\hat p_{_{t}})d\tilde
r_{_{A}}+\frac{T_{_{H}}}{G}\pi\tilde r_{_{A}}^{2}dF.
\end{equation}
The Misner Sharp energy $E$ in general relativity is
defined as $E\equiv\tilde r_{_{A}}/2G$. Since $G_{_{eff}}=G/F$
in $F(\tilde R)$ gravity, may be
written as
\begin{equation}\label{25}
\hat E=\frac{\tilde r_{_{A}}F}{2G},
\end{equation}
by combining Eqs. (\ref{16}) and (\ref{25}), we get
\begin{equation}\label{26}
\hat E=\frac{3FH^{2}}{8\pi G}V=\frac{1}{3\lambda-1}V\hat\rho_{_{t}}.
\end{equation}
Where $V=4\pi\tilde r_{_{A}}^{3}/3$ is the volume inside the
apparent horizon. It shows that from the Eq. (\ref{26}) $\hat E$
corresponds to the total intrinsic energy. It is also clear that
from this Eq. (\ref{26}) that $F\geq0$ so that $\hat E\geq0$. The
effective gravitational coupling in $F(\tilde R)$ gravity becomes
positive (no ghost) .
.

Using Eqs. (\ref{13}) and (\ref{14}), we have
\begin{equation}\label{27}
d\hat E=-\frac{4\pi}{3\lambda-1}\tilde r_{_{A}}^{3}H
(\hat\rho_{_{t}}+\hat p_{_{t}})dt +\frac{4\pi}{3\lambda-1}\tilde
r_{_{A}}^{2}\hat\rho_{_{t}}d\tilde r_{_{A}}+\frac{\tilde
r_{_{A}}}{2G}dF.
\end{equation}
By using the Eqs. (\ref{24}) and (\ref{27})
\begin{equation}\label{28}
T_{_{H}}d\hat S=-d\hat E+\frac{2\pi}{3\lambda-1}\tilde
r_{_{A}}^{2}(\hat\rho_{_{t}}-\hat p_{_{t}})d\tilde
r_{_{A}}+\frac{\tilde r_{_{A}}}{2G}(1+2\pi\tilde
r_{_{A}}T_{_{H}})dF.
\end{equation}
By introducing the work density \cite{Bardeen},
\begin{eqnarray}\label{29}
\hat{W}&\equiv&-\frac{1}{2}(T^{(M)\alpha\beta}h_{_{\alpha\beta}}
+\hat{T}^{(DE)\alpha\beta}h_{_{\alpha\beta}}),\\\label
{30}&=&-\frac{1}{2}(\hat{\rho}_{_{t}}-\hat{\rho}_{_{t}}).
\end{eqnarray}
With $\hat T^{(de)\alpha\beta}$ being the energy-momentum tensor of
the dark components, Eq. (\ref{29}) is rewritten as
\begin{equation}\label{31}
T_{_{H}}d\hat S=-d\hat E+\frac{1}{3\lambda-1}\hat W dV+\frac{\tilde
r_{_{A}}}{2G}(1+2\pi\tilde r_{_{A}}T_{_{H}})dF.
\end{equation}
Which can be described as
\begin{equation}\label{32}
T_{_{H}}d\hat S+T_{_{H}}d_{_{i}}\hat S=-d\hat
E+\frac{1}{3\lambda-1}\hat WdV.
\end{equation}
Where
\begin{eqnarray}\nonumber
d_{_{i}}\hat S&=&-\frac{1}{T_{_{H}}}\frac{\tilde
r_{_{A}}}{2G}(1+2\pi\tilde r_{_{A}}T_{_{H}})dF= -(\frac{\hat
E}{T_{_{H}}}+\hat S)\frac{dF}{F}\\\label{33}
&=&-\frac{\pi}{GH^{2}}\frac{4H^{2}+\dot{H}}{2H^{2}+\dot{H}}dF.
\end{eqnarray}
The term $d_{_{i}}\hat S$ is a additional term which can be
interpreted as an entropy production term in the non-equilibrium
thermodynamics.
\subsection{Second Law of Thermodynamics}.

   Recently, the second law of thermodynamics has been studied in the
context of modified $F(R)$ Ho\v{r}ava-Lifshitz gravitational theory. It
may be interesting to investigate its validity in $F(\tilde R)$
gravity. For this purpose, we have to show that
\begin{equation}\label{34}
\Xi\equiv\frac{d\hat S}{dt}+\frac{d_{_{i}}\hat S}{dt}+ \frac{d\hat
S_{_{t}}}{dt}\geq0.
\end{equation}
Where $\hat S$ is the horizon entropy in $F(\tilde R)$ gravity and
$\hat S_{_{tot}}$ is the entropy due to all the matter and energy
sources inside the horizon. The Gibbs equation including all matter
and energy fluid is given by
\begin{equation}\label{35}
T_{_{H}}dS_{_{t}}=d(\rho_{_{t}}V)+p_{_{t}}dV=Vd\rho_{_{t}}+
(\rho_{_{t}}+p_{_{t}})dV.
\end{equation}
Where $T_{_{H}}$ and $\hat S_{_{t}}$ denotes the temperature and
entropy of total energy inside the horizon, respectively. The main assumption is that here we suppose that inside and outside of the apparent horizon remain in thermal equilibrium with the same temperature.

 By using the Eqs. (\ref{8}), (\ref{32}) and (\ref{35}), we obtain,
\begin{equation}\label{36}
\Xi=\frac{F}{2G}\frac{\dot{H}^{2}}{H^{4}}.
\end{equation}
\begin{equation}\label{37}
J=144H^{2}\dot{H}^{2}F\geq0.
\end{equation}
Which is always met because $F>0$ and $\hat{E}>0$. Hence the second
law of thermodynamics can be satisfied in $F(R)$ Ho\v{r}ava-Lifshitz
gravity. We can see that from Eq. (\ref{37}) $J\geq0$ irrespective
of the sign of $\dot{H}$.

We conclude that we have used the Physical temperature as the
temperature of apparent horizon. This temperature clearly depends on
the energy momentum tensor of the dark components of $F(R)$
Ho\v{r}ava-Lifshitz gravity. The temperature of matter species in a
cosmological setup is determined in a standard way. We have
concentrated in our discussions (second law of thermodynamics in
$F(R)$ Ho\v{r}ava-Lifshitz gravity) on the case in which temperature of
the universe inside the horizon is equal to that of the apparent
horizon.
\section{Equilibrium Description of Thermodynamics in
$F(R)$ Ho\v{r}ava-Lifshitz Gravity} \label{equil}
In the case of non-equilibrium
description of thermodynamics the entropy production term
$d_{_{i}}\hat S$, the R.H.S of Eq. (\ref{13}) does not vanish and
the equation of continuity for $\hat\rho_{_{de}}$ and $\hat
P_{_{de}}$ does not hold for this purpose. We demonstrated that in
the case of equilibrium description of thermodynamics by redefining
the energy density and pressure of dark components to meet the
continuity equation. So, there can be no extra entropy production
term in the equilibrium description in $F(R)$ Ho\v{r}ava-Lifshitz
gravity.
\subsection{Energy Density and Pressure of Dark
Components} The Friedmann equations are for equilibrium description
in $F(R)$ Ho\v{r}ava-Lifshitz gravity
\begin{equation}\label{38}
H^{2}=\frac{\kappa^{2}}{3(3\lambda-1)}\rho_{_{eff}},\quad
\dot{H}=\frac{-\kappa^{2}
}{2(3\lambda-1)}(\rho_{_{eff}}+p_{_{eff}}),
\end{equation}
\begin{equation}\label{39}
\rho_{_{eff}}=\rho_{_{de}}+\rho_{_{m}}.
\end{equation}
\begin{equation}\label{40}
p_{_{eff}}=p_{_{de}}+p_{_{m}}.
\end{equation}
The energy density and pressure of dark components can be rewritten
as
\begin{eqnarray}\nonumber
\rho_{_{de}}&=&-f(\tilde R)+{6}(1-3\lambda+3\mu)H^{2}F(\tilde
R)+{6}{\mu}{\dot{H}F(\tilde{R})}{6}\\\label{41}&\times&{\mu}H\frac{d
F(\tilde{R})}{dt}.
\end{eqnarray}
\begin{eqnarray}\nonumber
p_{_{de}}&=&f(\tilde R)-{2}(1-3\lambda+3\mu)(3H^{2}+\dot{H})F(\tilde
R)-{2}\\\label{42}&\times&(1-3\lambda)H\frac{dF(\tilde
R)}{dt}+{2}{\mu}\frac{d^{2}F(\tilde R)}{dt^{2}}.
\end{eqnarray}
Which clearly satisfy the standard equations of continuity, i.e,
\begin{equation}\label{43}
\dot\rho_{_{de}}+3H(\rho_{_{de}}+p_{_{de}})=0.
\end{equation}
\subsection{First law of thermodynamics}
By using Eq. (\ref{38}) and (\ref{17}), we get
\begin{equation}\label{44}
\frac{1}{4\pi G}d\tilde
r_{_{A}}=\frac{r_{_{A}}^{3}H}{3\lambda-1}(\rho_{_{t}}+p_{_{t}})dt.
\end{equation}
Where $\rho_{_{t}}=\rho_{_{de}}+\rho_{_{m}},
p_{_{t}}=p_{_{de}}+p_{_{m}}$, by introducing the horizon entropy
$S=A/4G$ and using Eq. (\ref{44}), we have
\begin{equation}\label{45}
\frac{1}{2\pi\tilde r_{_{A}}}dS=\frac{4\pi}{3\lambda-1}\tilde
r_{_{A}}^{3}H(\rho_{_{t}}+p_{_{t}})dt.
\end{equation}
From the horizon temperature in Eq. (\ref{23}) and (\ref{45}) , we
get,
\begin{equation}\label{46}
T_{_{H}}d{S}=\frac{4\pi}{3\lambda-1}\tilde
r_{_{A}}^{3}H(\rho_{_{t}}+p_{_{t}})dt-\frac{2\pi}{3\lambda-1} \tilde
r_{_{A}}^{2}(\rho_{_{t}}+p_{_{t}})d\tilde r_{_{A}}.
\end{equation}
By defining the Misner-sharp energy as
$E=\frac{\tilde r_{_{A}}}{2G}$.
By solving the Eqs. we have,
\begin{equation}\label{47}
E=\frac{3FH^{2}}{8\pi G}V=\frac{1}{3\lambda-1}V\rho_{_{t}}.
\end{equation}
We get
\begin{equation}\label{48}
dE=-\frac{4\pi}{3\lambda-1}\tilde r_{_{A}}^{3}H(\rho_{_{t}}
+p_{_{t}})dt+\frac{4\pi}{3\lambda-1}\tilde
r_{_{A}}^{2}\rho_{_{t}}d\tilde r_{_{A}}.
\end{equation}
It is noted that, there does not exists any additional term
proportional to the $dF$ on the R.H.S due to the continuity equation
Eq (\ref{44}). From Eqs. (\ref{46}) and (\ref{48}), we get the
following equation corresponding to the first law of thermodynamics.
\begin{equation}\label{49}
T_{_{H}}dS=-dE+\frac{1}{3\lambda-1}W dV.
\end{equation}
Where the work density is given by
\begin{equation}\label{50}
W=\frac{1}{2}(\rho_{_{t}}-p_{_{t}}).
\end{equation}
So, by redefining $\rho_{_{DE}}$ and $P_{_{DE}}$, the equation of
continuity can be satisfies, we can realize the existence of the equilibrium
thermodynamical phase in $f\tilde{(R)}$ gravity.

by using the Eqs. (\ref{38}), (\ref{45}), and (\ref{43}), we get
\begin{equation}\label{50}
\dot S=-\frac{2\pi}{G}\frac{\dot H}{H^{3}}.
\end{equation}
Since $\dot S\propto-\dot H/H^{3}$, the horizon entropy increases in
the expanding universe as long as the null energy condition
$\rho_{_{t}}+P_{_{t}}\geq0$ is satisfied, in which $\dot H\leq0$.

There are two main reasons why we can obtain the equilibrium
description of thermodynamics:
\begin{itemize}
\item First of all, the Bekenstein-Hawking area entropy is valid here just by a formal redefinition of the effective $G_{eff}$.

\item We satisfy continuty equation by redefining of the effective
energy density and
pressure of dark components .
\end{itemize}
The basic relation between
the horizon entropy $S$ in
the equilibrium description and $\hat S$ in the non-equilibrium
description are given as follows:
\begin{equation}\label{51}
dS=d\hat S+d_{_{i}}\hat S+\frac{\tilde
r_{_{A}}}{2GT_{_{H}}}dF-\frac{2\pi(1-F)}{G}\frac{\dot H}{H^{3}}dt.
\end{equation}
By using the relation (\ref{33}), (\ref{44}) and (\ref{51}), we
have,
\begin{equation}\label{52}
dS=\frac{1}{F}d\hat S+\frac{1}{F}\frac{2H^{2}+\dot H}{4H^{2} +\dot
H}d_{_{i}}\hat S.
\end{equation}
Where $d_{_{i}}\hat S$  is given by Eq. (\ref{33}). Because of $dF\neq0$, we obtain a non zero difference
between $S$ and $\hat S$.
\subsection{Second Law of Thermodynamics} In the case of
the equilibrium description, to evaluate the second law of
thermodynamics, we write the Gibbs equation in terms of all matter
and energy fluid as
\begin{equation}\label{F53}
T_{_{H}}dS_{_{t}}=d(\rho_{_{t}}V)+p_{_{t}}dV=V
d\rho_{_{t}}+(\rho_{_{t}}+p_{_{t}})dV.
\end{equation}
The second law of thermodynamics argues that the total entropy of the system never decreases in time:
\begin{equation}\label{F54}
\frac{dS_{_{sum}}}{dt}\equiv\frac{dS}{dt}+\frac{{dS}_{_{t}}}{dt}\geq0.
\end{equation}
Where $S_{_{sum}}=S+S_{_{t}}$. Consequently, we have
\begin{equation}\label{F55}
\frac{{dS}_{_{sum}}}{dt}=\frac{2\pi}{G}\frac{\dot
H^{2}}{H^{2}}\frac{1}{H(2H^{2}+\dot H)}.
\end{equation}
By using $V=4\pi\tilde r_{_{A}}^{3}/3$, and Eqs.
(\ref{23})(\ref{38}) and (\ref{50}). Hence the relation with Eq.
(\ref{F54}) leads to the condition
\begin{equation}\label{F56}
Y\equiv12H(2H^{2}+\dot H)\geq0.
\end{equation}
In the flat FLRW expanding background $(H\geq0)$, the second law of
thermodynamics can be satisfied in $F(R)$ Ho\v{r}ava-Lifshitz gravity,
$R=6(2H^{2}+\dot H)$  for the flat FLRW space-time and the
condition in Eq.(\ref{F56}) clearly holds as $\tilde R\geq0$.

\section{Summary and conclusion} Thermodynamic laws in the non relativistic regime of $f(R)$ gravity,
$F(R)$ Ho\v{r}ava-Lifshitz gravity investigated in both equilibrium and non equilibrium modes. By assuming that the inside and outside of apparent horizon are in thermal equilibrium, we proved that the second law of thermodynamics can be
satisfied for both cases. In the non-equilibrium framework, it has
been shown that the second law of the thermodynamics can be satisfied
regardless of the sign of the time derivative of the Hubble
parameter and in the equilibrium framework, the second law of
thermodynamics can be shown by analogy with the same non-negative
quantity which is related to the scalar curvature in (GR) is
positive or equal to zero in cosmology.

Finally, we can conclude that our result of second law of
thermodynamics in  $F(R)$ Ho\v{r}ava-Lifshitz gravity is non-trivial.

\end{document}